\documentclass[reprint,superscriptaddress]{revtex4-1}
\usepackage[usenames,dvipsnames,svgnames,table]{xcolor}

\usepackage{float}
\usepackage{amsmath}
\usepackage{amssymb}
\usepackage{graphicx}
\usepackage{hyperref}
\usepackage{booktabs}
\usepackage{pstricks-add}
\newcommand{\changes}[1]{{\color{black} #1}} 

\usepackage[caption=false]{subfig}

\makeatother

\begin{document}

\title{Transient structures in rupturing thin-films: Marangoni-induced symmetry-breaking pattern formation in viscous fluids}

\author{Li \surname{Shen}}
\affiliation{Department of Mechanical Engineering, Imperial College London, London,
SW7 2AZ, United Kingdom}
\thanks{Corresponding Author}
\email{l.shen14@imperial.ac.uk}

\author{Fabian Denner}
\affiliation{Lehrstuhl f\"{u}r Mechanische Verfahrenstechnik, Otto-von-Guericke-Universit\"{a}t Magdeburg, Universit\"atsplatz 2, 39106 Magdeburg, Germany}
\author{Neal Morgan}
\affiliation{Shell Global Solutions Ltd, Shell Centre, York Road, London, SE1 7NA, UK}
\author{Berend van Wachem}
\affiliation{Lehrstuhl f\"{u}r Mechanische Verfahrenstechnik, Otto-von-Guericke-Universit\"{a}t Magdeburg, Universit\"atsplatz 2, 39106 Magdeburg, Germany}
\author{Daniele Dini}
\affiliation{Department of Mechanical Engineering, Imperial College London, London,
SW7 2AZ, United Kingdom}

\date{\today}
\begin{abstract}

In the minutes immediately preceeding the rupture of a soap bubble, distinctive and repeatable patterns can be observed. These quasi-stable transient structures are associated with the instabilities
of the complex Marangoni flows on the curved thin film in the presence
of a surfactant solution. Here, we report a generalised Cahn-Hilliard-Swift-Hohenberg
model derived using asymptotic theory which describes the quasi-elastic
wrinkling pattern formation and the consequent coarsening dynamics
in a curved surfactant-laden thin film. By testing the theory against
experiments on soap bubbles, we find quantitative agreement with the
analytical predictions of the nucleation and the early coarsening
phases associated with the patterns. Our findings provide fundamental physical understanding that can be used to (de-)stabilise thin films in the presence of surfactants and have important implications for both natural and industrial contexts, such as the production of thin coating films, foams, emulsions and sprays. 

\end{abstract}
\maketitle

The Marangoni effect inherently produces non-linear structures within
fluid flow, the formation of these structures will inevitably lead
to symmetry breaking within the system. The premise of symmetry breaking
is present in a wide range of phenomena, not least in the interfacial
fluid context; from the Higgs mechanism \cite{Higgs1964} in particle physics, to solid
crystallisation \cite{Onuki2002} and to the functions of the cell structure \cite{Palmer2004}. The analysis of these symmetry-breaking phenomena typically
involve a reduction of the complex system into field variables, which
then form an effective field theory. The Ginzburg-Landau theory of
phase transitions \cite{Ginzburg1950}, the Cahn-Hilliard (CH) of phase separation \cite{Cahn1958}, the Swift-Hohenberg (SH) of convective instability \cite{Swift1977} and the Turing
description of reaction-diffusion patterns \cite{Turing1990} are some examples of this
approach which identifies local and generic structures within the
system. However, the systematic derivation of nonlinear field theories
remains challenging; not only is reproducing macroscopic behaviours from microscopic
patterns difficult, but bridging the gap between macro-
and microscopic dynamics is also non-trivial. Instead, abstract symmetries
and bifurcation methods are often used to derive the effective field
equations with a large number of undetermined parameters \cite{Marsden2003}, which limit
comparisons with experimental data. Nonetheless, under certain limits
of hydrodynamic motion they remain effective at predicting the formation
of local structures which typify a certain configuration of field
variables. 

In the absence of fluid flow, quasi-static patterns can form on thin films of condensed materials, as can be seen through the buckling and wrinkling of soft elastic membranes \cite{Stoop2015}. There are numerous biological examples of these patterns, e.g. the cortical convolutions of the brain \cite{Tallinen2016}, fingerprint \cite{Cerda2003} and skin wrinkling \cite{Efimenko2005}. The presence of curvature in biological thin film wrinkling systems, together with applied stresses can induce morphogenesis \cite{Cao2008} and indeed scalar field equations can be derived in cases of quasi-static morphogenesis on soft elastic membranes \cite{Stoop2015} to predict transitions and symmetry-breaking patterns quantitatively. This thus enables targeted engineering of pattern formation on nanosurfaces \cite{Bowden1998} and advances in microlens array fabrication \cite{Chan2006}. In the presence of fluid flow, the governing Navier-Stokes equation and the associated pattern formation dynamics admit complex nonlinearities (Supplementary Video 1).

Here we systematically derive and experimentally test
an effective asymptotic field theory which predicts quantitatively
the surface-pattern formation in a curved thin film fluid system embued
with surface-active materials. Using asymptotic expansion methods under the thin-film lubrication approximation allows for a detailed quantitative analysis of the morphogenesis process as well as predictions of symmetry-breaking and transitional behaviours. A generalised fourth-order hybrid Swift-Hohenberg-Cahn-Hilliard (SHCH) equation for the normal displacement field of an arbitrarily curved thin liquid film is derived to the leading order. 

\begin{figure*} 
	  \includegraphics[width=17cm]{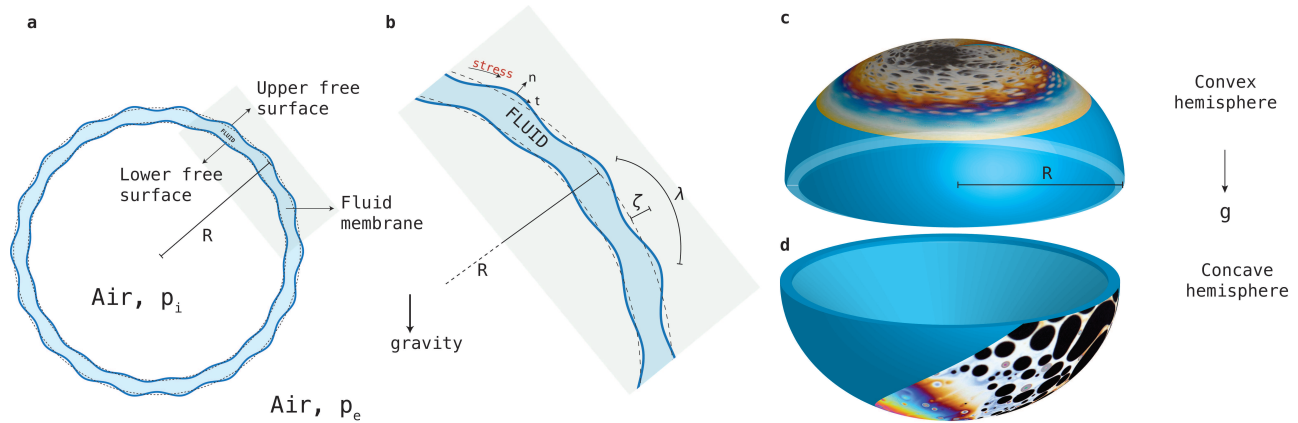}
  \caption{Notation and experimental system. \textbf{a}. Schematic of a curved spherical thin film of radius $R$ and film thickness $2h$. The air pressure difference between the interior and the exterior of the bubble is given by $\Delta p=p_e-p_i$ and $\mathrm{g}$ is the gravitational acceleration. \textbf{b}. The film is driven towards an instability pattern with wavelength $\lambda$ with height displacement $\zeta$ in the radial direction. \textbf{c-d}. Convex and concave hemispherical geometries with the experimental images overlayed at the position where they are observed.}
  \label{fig:schematic}
\end{figure*}

Due to the transient nature of the morphogenic process (see Supplementary Video 2), in the long-term the system can be observed \cite{Shen2017a} to tend to either a SH-like hexagonal dot or labyrinth-patterned stable state, similar to the quasi-static \cite{Stoop2015} case, or alternatively tending towards a black Newtonian film after undergoing a CH-like coarsening process. In the SH-dominant regime, a weakly-nonlinear analysis yields a stability criteria for the steady state patterns. In the CH-dominant regime, fractal dimensional analysis on the simulation results provide a direct quantitative agreement with experiments of pattern formation on curved soap films and bubbles. In both regimes, a useful feature of the asymptotic expansion is the ease of extension to arbitrarily higher orders and therefore provides a natural framework for future studies of similar transient morphogenic phenomena in a thin liquid film in bio-physical systems.

\section*{Theory of pattern formation on thin-films}

Starting from the time-dependent Stokes' equation under thin-film
conditions, i.e. film thickness $h\ll1$, we obtain the amplitude
equation through an asymptotic expansion in the limit $\epsilon\rightarrow0$,
where $\epsilon$ is the non-dimensional distance to the critical
excitation wavenumber. In the presence of the fluid flow, the resulting
amplitude equation derived under the assumption of quasi-static flow offers an insight into the transient
pattern formation processes we observe experimentally.

To match our experimental conditions, which are shown in figure (\ref{fig:schematic}), we
consider two cases: a convex and a concave hemispherical geometry as in figure (\ref{fig:schematic}c-d)
with $R/h\gg1$, where $R$ is the radius of the hemisphere.  Decomposing the liquid surface, we consider a wave-form $
F(x,y,z,t)=z-\zeta(x,y,t),$ where $\zeta(t)$ is the time-dependent wave amplitude which satisfies
the small-amplitude condition that $\zeta\ll\lambda=2\pi/k$ and $\mathrm{d}\zeta/\mathrm{d} t \ll v_{c}=\omega/k$,
where $\lambda,k,\omega,v_{c}$ are the wavelength, wavenumber, frequency
and the phase velocity, respectively. Assuming an overall overdamped region for $\omega$, i.e. $\mathrm{Re}(\omega)=0$, we obtain at the leading order the amplitude
equation 
 \begin{alignat}{1}
\partial_{\tau}\zeta & =\Delta_{2}\left(c_{1}\zeta+c_{2}\zeta^{2}+c_{3}\zeta^{3}+c_{4}\Delta_{2}\zeta\right)\nonumber \\
 & \phantom{=}+c_{5}\zeta+c_{6}\zeta^{2}+c_{7}\zeta^{3},\label{eq:amp-eqn}
\end{alignat}  
where $\Delta_{2}$ is the 2-dimensional Laplacian operator, $\tau$
is the non-dimensional time, $c_{i}\in\mathbb{R}$ for $i=1,\ldots,7$
are coefficients which depend on the Laplace number $\mathrm{La}=\rho\sigma h_0/\mu^2$ and the velocity-independent Marangoni number $\mathrm{M}=\rho(\Delta \sigma)h_0/\mu^2$, where $\rho,\sigma,h_0,\mu$ are fluid density, surface tension coefficient, reference film thickness and dynamic viscosity, respectively. The $(c_{1},c_{2},c_{3})$-terms describe the nucleation and the coarsening processes of the
instability, typical of the CH-type equation, the $(c_{5},c_{6},c_{7})$-terms describe dynamics affected by convection, usually present in
the SH-type equation, and the $c_4$ bi-Laplacian term is common to both type of equations. Due to the fluid nature of the system, obtaining the coefficients $c_{i}$ for $i=1,\ldots,7$ is a non-trivial process. However, under the quasi-static assumption which is valid near the onset of instability and corresponds with the experimental cases under consideration, we can obtain the coefficients $c_1,\ldots,c_7$ analytically using the lubrication approximation in the thin-film fluid dynamics.  





The detailed derivation of the asymptotic expansion under the limit $\epsilon\rightarrow0$ can be found in the Supplementary Material, combined with the systematic analysis which enables us to express the coefficients in equation (\ref{eq:amp-eqn}) in terms of the non-dimensional parameters $\mathrm{La}$ and $\mathrm{M}$, which denote the effect of curvature and the localised strength of the gradient of the surface tension coefficient, respectively. 
The surface tension near the onset of instability for the initial time is assumed to follow the linear Henry isotherm $\sigma=\sigma_0-\alpha \Gamma$, where $\alpha=-|\partial\sigma/\partial\Gamma|$ for surfactant concentration $\Gamma$ and initial surface tension $\sigma_0$. 

In absence of surfactants, the critical damping wavelength, where the wave transitions from an underdamped to an overdamped regime occurs, is given by $\lambda_{\mathrm{c}}^{\mathrm{w}}=2\pi l_\mathrm{vc}/\epsilon^{\star 2}$ \cite{Shen2018}, where $l_\mathrm{vc}=\mu^2/(\rho \sigma)$ is the viscocapillary lengthscale and $\epsilon^\star$ is the largest positive root \cite{Shen2018} of $\mathrm{f}(\epsilon)=11\epsilon^6-18\epsilon^4-\epsilon^2-1$ . For soap films under consideration, the presence of the surfactant solution damps the surface waves of the system \cite{Shen2017} and thus propels the value of $\lambda_\mathrm{c}^{\mathrm{w}}$ to within the range of the interferometry techniques employed to measure the thickness of thin films and, indeed, the experimentally observed film thickness in the current study. 

Since the existence of real coefficients in the amplitude equation (\ref{eq:amp-eqn}) is a necessary condition for the instability dynamics to manifest itself physically, 
 this suggests the pattern-forming region overlaps with the overdamped capillary wave regime. In this overdamped regime, we require $\mathrm{i}\omega\in \mathbb{R}$ where the general complex frequency $\omega=\omega_1+\mathrm{i}\omega_2\in\mathrm{i}\mathbb{R}$ appears in the dispersion relation of a viscous system with non-trivial surface tension and gravity 
\cite{Harden1991,Jackle1999,Delgado2008a,Denner2016b,Shen2018} given by 
\begin{equation}
\omega_0^2+(\mathrm{i}\omega+2\nu k^{2})^{2}-4\nu^{2}k^{4}\left(1+\frac{\mathrm{i}\omega}{\nu k^{2}}\right)^{1/2}=0,
\end{equation}
where $\omega_0^2 = \sigma k^3/\rho+\mathrm{g}k$. For regions where $\mathrm{i}\omega\in \mathbb{R}$, the viscous effects of magnitude $\nu k^2$ dominates the restoring effects of the surface tension  $(\sigma k^3/\rho)^{1/2}$ and, hence, surface fluctuations are found \cite{Delgado2008a} to be exponentially damped  with $\omega_2\sim k$ and $\omega_1=0$. 

\begin{figure*}
  \includegraphics[width=17cm]{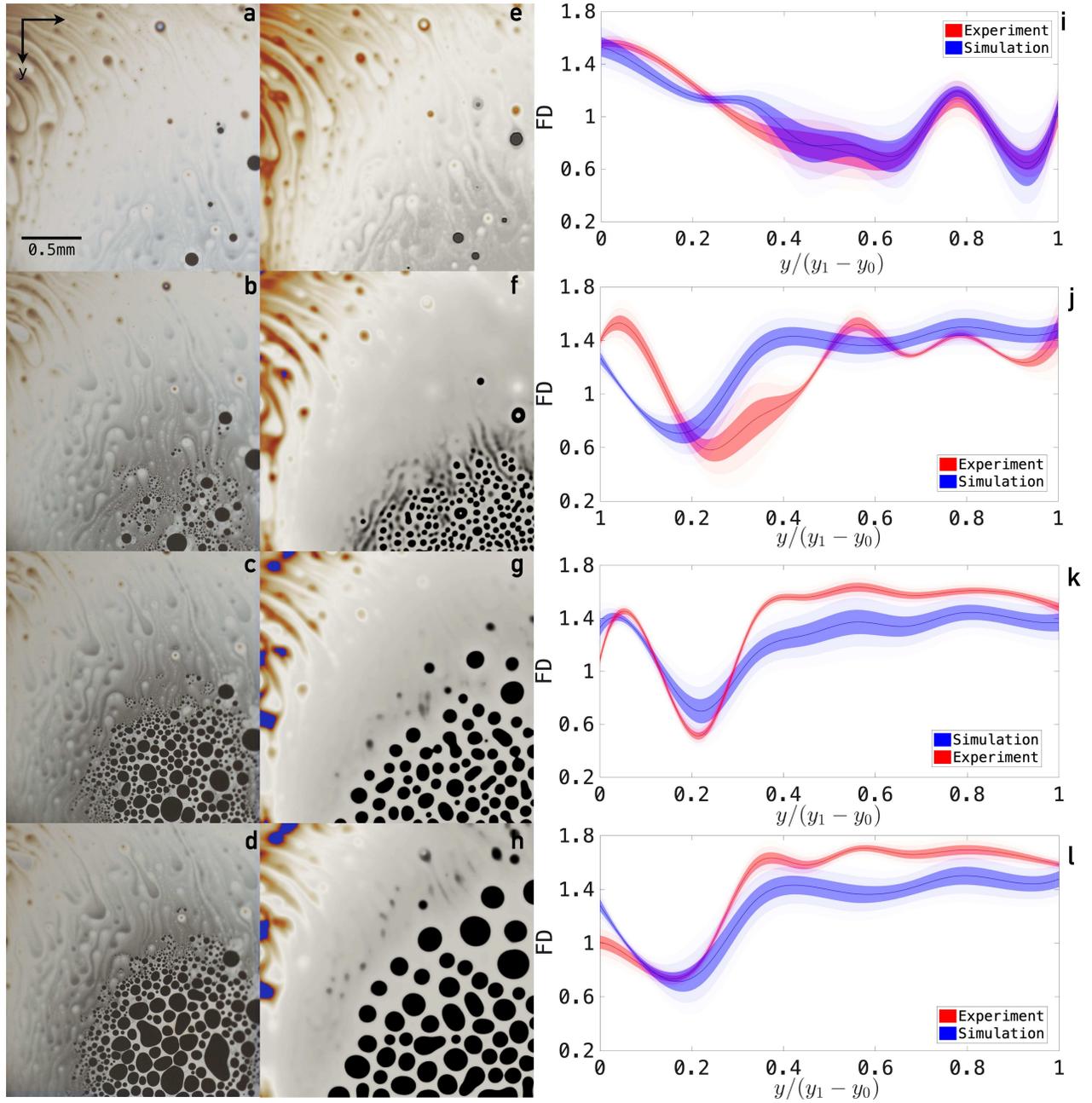}
  \caption{Experimental images $\mathbf{a}$-$\mathbf{d}$ and simulation results $\mathbf{e}$-$\mathbf{h}$ of pattern formation on a concave thin film near rupture. Subfigures $\mathbf{i}$-$\mathbf{l}$ show the numerical fractal dimensions of subfigures a-h where the $y_0$ and $y_1$ are the $y$ values for the top and bottom edge, respectively. The contours around the curve denote the standard deviations (sd) of the fractal dimension, with the deepest colour showing the bounds of $\tfrac{1}{2}\text{sd}$ and the faded colour giving the $1\text{sd}$ bounds.}
  \label{fig:patternformation}
\end{figure*}

Under the overdamped capillary wave regime assumption for the dynamics of the pattern formation, we conjecture that the wavelength of the pattern-forming instability $\lambda$ is proportional to the critical damping wavelength $\lambda_{\mathrm{c}}^{\mathrm{w}}$. This provides the relevant lengthscale for the critical wavelength $\lambda_\mathrm{c}$ at the inception of the pattern-forming instability, with $\lambda \sim\lambda_\mathrm{c}^{\mathrm{w}}$. 
To relate the pattern-forming wavelength with the normal surface displacement amplitude, we consider that under the quasi-static assumption where the fluid velocity lengthscale $U\ll1$ and using the symmetry of the system, the boundness condition (see Supplementary Material) on the vertical velocity $w$ as the film height $z\rightarrow0$ leads to the power-law relation 
\begin{equation}
\bar{\lambda} \sim \bar{\zeta}^{2/3},	\label{eq:th-wa-relation}
\end{equation}
where the non-dimensional normal displacement amplitude and the wavelength of the pattern-forming instability are given by  $\bar{\zeta}=\zeta/\zeta_0$ and $\bar{\lambda}=\lambda/\lambda_0$, respectively, where $\zeta_0$ is the reference normal displacement amplitude and $\lambda_0$ is the initial wavelength. This relation links $\zeta$ and $\lambda$, which suggests that results from the experimental thin-film interferometry technique can be used directly to probe the lengthscale of the instability wavelength. 

Finally, we derive the real coefficients in the amplitude equation in the Supplementary Material, which is shown in table (1). These coefficients together with equation (\ref{eq:amp-eqn}) give a complete leading-order description of the transient pattern-forming dynamics in the overdamped capillary wave regime under quasi-static fluid flow. 

\begin{table}
\vspace{0.25cm}
\begin{centering}
\begin{tabular}{l}
\addlinespace
$c_{1}=\tfrac{2}{15}\hat{\mathrm{M}}-\hat{\mathrm{L}}$\tabularnewline\addlinespace
\addlinespace
\addlinespace
$c_{2}=-\frac{3}{4}\hat{\mathrm{L}}$\tabularnewline\addlinespace
\addlinespace
\addlinespace
$c_{3}=\frac{117}{80}-\hat{\mathrm{L}}$\tabularnewline\addlinespace
\addlinespace
\addlinespace
$c_{4}=-\frac{38}{75}+\frac{6}{15}\hat{\mathrm{L}}$\tabularnewline\addlinespace
\addlinespace
\addlinespace
$c_{5}=\frac{1}{3}\hat{\mathrm{M}}$\tabularnewline\addlinespace
\addlinespace
\addlinespace
$c_{6}=\frac{1}{4}\hat{\mathrm{M}}$\tabularnewline\addlinespace
\addlinespace
\addlinespace
$c_{7}=\tfrac{19}{45}\hat{\mathrm{M}}+\tfrac{2}{5}\hat{\mathrm{L}}-\tfrac{28}{75}$\tabularnewline\addlinespace
\addlinespace
\end{tabular}
\par\end{centering}
\caption{List of normalised parameters for equation (\ref{eq:amp-eqn}), where $\hat{\mathrm{L}}=\mathrm{La}^{-2}$ and $\hat{\mathrm{M}}=\mathrm{M}^{-2}$}
\label{tab:coefficients}
\end{table}

\section*{Stability and Hysteresis}

The pattern selection and coarsening at the onset of the instability are nonlinear processes and thus cannot be derived by a linear stability analysis. Furthermore, the pattern instability is highly transient and numerical simulations of equation (\ref{eq:amp-eqn}) in the parameter space of $\mathrm{M}$ and $\mathrm{La}$ do not yield steady states, but transitions from the hexagonal dots to the labyrinth phases or the intermediate mixture phase where hexagonal dots and labyrinths coexist. The transition can be understood to be the consequence of terms with even parity in equation (\ref{eq:amp-eqn}) breaking the invariance of the solution under the reflection map $\Theta(\zeta)= -\zeta$. The coefficients of the symmetry-breaking terms, under the overdamped thin-film approximation, are shown in table (\ref{tab:coefficients}) to depend on $\mathrm{La}$ and $\mathrm{M}$, and thus we can expect a role for both curvature and Marangoni effect in the symmetry-breaking transition. In similar symmetry-breaking behaviour in the classical SH-type systems \cite{Cross1993,Golovin2006,Stoop2015} where the inclusion of similar terms destroy the $\Theta$-invariance of the solution, a qualitative comparison suggests small values of the curvature $\mathrm{La}$ and the surface tension gradient $\mathrm{M}$ would result in a hexagonal dot phase, while larger values of $\mathrm{La}$ and $\mathrm{M}$ would yield a labyrinth phase. 

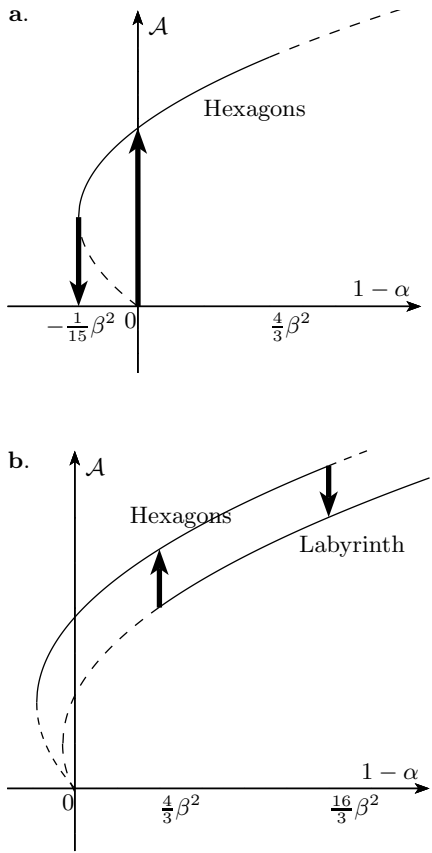
\begin{figure}
\vspace{1cm}
\begin{centering}
\psset{xunit=1.75cm,yunit=1.25cm,algebraic=true,dimen=middle,dotstyle=o,dotsize=3pt 0,linewidth=0.5pt,arrowsize=3pt 2,arrowinset=0.25}
\begin{pspicture*}(-0.989608812939,-0.710719308379)(2.15267011823,3.15074313348)
\psaxes[labelFontSize=\scriptstyle,xAxis=true,yAxis=true,labels=none,Dx=0.5,Dy=0.5,ticksize=0pt 0,subticks=2]{->}(0,0)(-0.989608812939,-0.710719308379)(2.15267011823,3.15074313348)[$1-\alpha$,140] [$\mathcal{A}$,-40]
\psplot[plotpoints=200]{-0.44999782120481285}{1}{sqrt((x-0.0500000000000)*2.00000000000+1.00000000000)+0.950000000000}
\psplot[linestyle=dashed,dash=4pt 4pt,plotpoints=200]{-0.44999782120481285}{0}{-sqrt((x-0.0500000000000)*2.00000000000+1.00000000000)+0.950000000000}
\psline[linewidth=2pt]{->}(0.00125,0.)(0.,1.89868329805)
\psline[linewidth=2pt]{->}(-0.45,0.95)(-0.45,0.)
\rput[tl](-0.7,-0.07){$-\frac{1}{15}\beta^2$}
\rput[tl](-0.102180127227,-0.07){$0$}
\rput[tl](1,-0.07){$\frac{4}{3}\beta^2$}
\rput[tl](0.5,2.2){Hexagons}
\rput[tl](-0.98,3.15){$\mathbf{a}.$}
\psplot[linestyle=dashed,dash=4pt 4pt,plotpoints=200]{1}{2.152670118229635}{sqrt((x-0.0500000000000)*2.00000000000+1.00000000000)+0.950000000000}
\end{pspicture*}
\vspace{1cm}

\psset{xunit=1.125cm,yunit=1.2cm,algebraic=true,dimen=middle,dotstyle=o,dotsize=3pt 0,linewidth=0.5pt,arrowsize=3pt 2,arrowinset=0.25}
\begin{pspicture*}(-0.799439590146,-0.693909297713)(4.19192076157,3.74464258189)
\psaxes[labelFontSize=\scriptstyle,xAxis=true,yAxis=true,labels=none,Dx=0.5,Dy=1.,ticksize=0pt 0,subticks=2]{->}(0,0)(-0.799439590146,-0.693909297713)(4.19192076157,3.74464258189)[$1-\alpha$,140] [$\mathcal{A}$,-40]
\psplot[plotpoints=500]{-0.44999944328291375}{3}{sqrt((x-0.0500000000000)*2.00000000000+1.00000000000)+0.950000000000}
\psplot[linestyle=dashed,dash=3pt 3pt,plotpoints=200]{-0.44999944328291375}{0}{-sqrt((x-0.0500000000000)*2.00000000000+1.00000000000)+0.950000000000}
\psplot[linestyle=dashed,dash=3pt 3pt,plotpoints=200]{3}{4}{sqrt((x-0.0500000000000)*2.00000000000+1.00000000000)+0.950000000000}
\rput[tl](-0.15,-0.07){$0$}
\rput[tl](1,-0.07){$\frac{4}{3}\beta^2$}
\rput[tl](3,-0.07){$\frac{16}{3}\beta^2$}
\psplot[plotpoints=500]{1}{4.191920761572669}{sqrt((x-0.360000000000)/0.500000000000+1.00000000000)+0.500000000000}
\psplot[linestyle=dashed,dash=4pt 4pt,plotpoints=200]{-0.14}{1}{sqrt((x-0.360000000000)/0.500000000000+1.00000000000)+0.500000000000}
\psplot[linestyle=dashed,dash=4pt 4pt,plotpoints=200]{-0.14469789600966343}{0}{-sqrt((x-0.355297580118)/0.500000000000+1.00000000000)+0.500125771005}
\rput[tl](0.65,3.12){Hexagons}
\rput[tl](2.65,2.8){Labyrinth}
\rput[tl](-0.79,3.74){$\mathbf{b}.$}
\psline[linewidth=2pt]{->}(1,2.00563276818)(1,2.6525853943)
\psline[linewidth=2pt]{->}(3,3.57850245162)(3,3.00770407649)
\end{pspicture*}

\par\end{centering}
  \caption{$\mathbf{a}$. The first hysteresis cycle for the amplitude equation near $1-\alpha=0$. The undisturbed state transforms into a hexagonal state at $1-\alpha=0$. $\mathbf{b}$. The second hysteresis cycle for the amplitude equation near $1-\alpha=\frac{4}{3}\beta^2$ and $1-\alpha=\frac{16}{3}\beta^2$ where the hexagonal state turns into the labyrinth state.}
  \label{fig:6hysteresis1}
\end{figure}

\begin{figure*}
  \includegraphics[width=18cm]{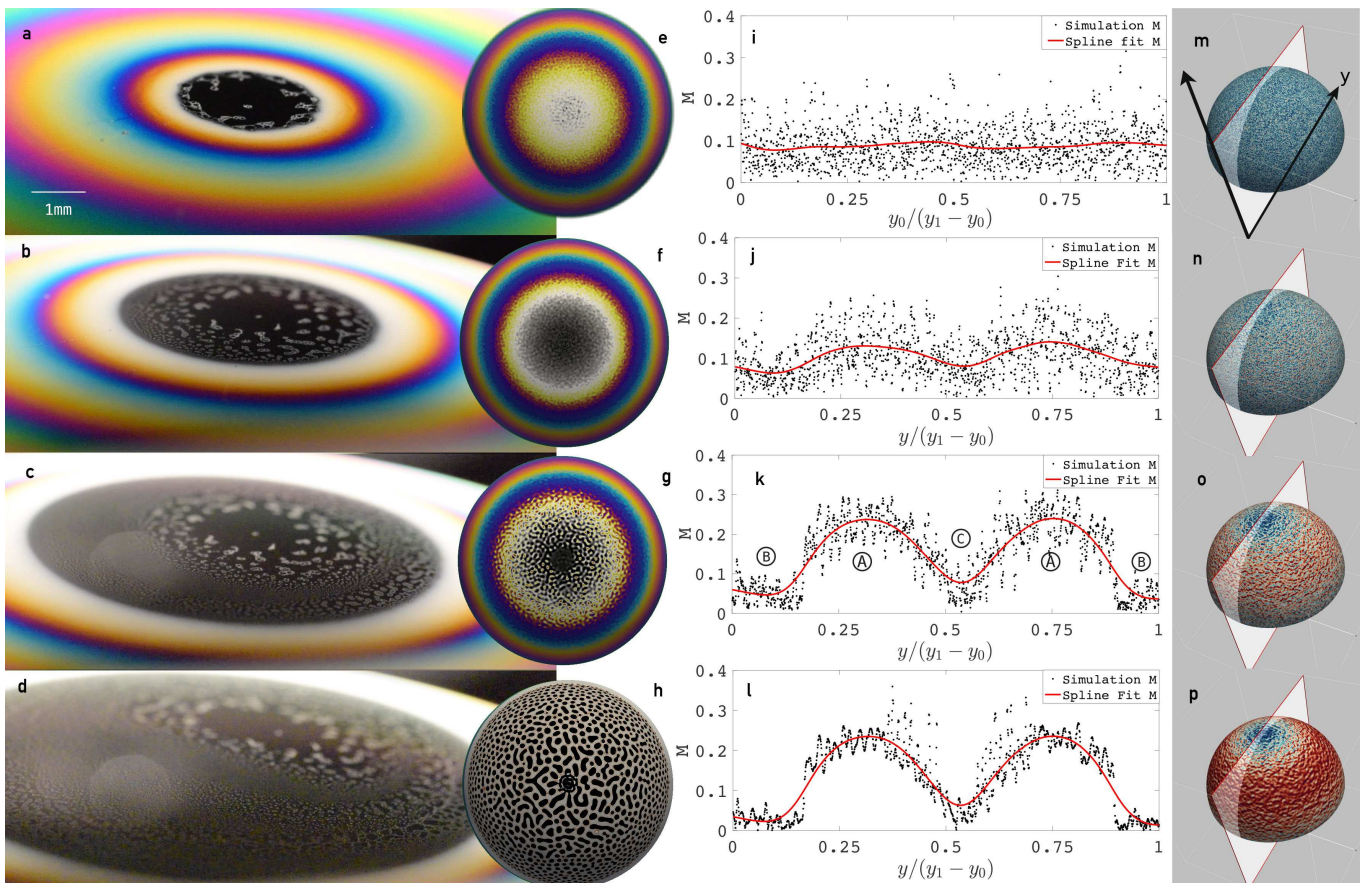}
  \caption{Experimental $\mathbf{a}$-$\mathbf{d}$ and simulation $\mathbf{e}$-$\mathbf{h}$ images of pattern formation on a curved thin film with associated $\mathbf{i}$-$\mathbf{l}$ plots of the Marangoni number and the Marangoni field $\mathbf{m}$-$\mathbf{p}$ on the hemisphere, where $y_0$ and $y_1$ are the left- and right-most $y$ values of the cross-section shown in $\mathbf{m}$.}
  \label{fig:hemispherepf}
\end{figure*}

The amplitude equation (\ref{eq:amp-eqn}) can be recast in the normal form of a mixed SHCH-type system as
\begin{equation}
\dot{\mathrm{A}}=-\Delta\frac{\mathrm{d}f(\mathrm{A})}{\mathrm{d}\mathrm{A}}+\Delta^{2}\mathrm{A}-\alpha\mathrm{A}-\beta\mathrm{A}^{2}-\mathrm{A}^{3},\label{eq:amplitude-equation-doublewell}
\end{equation}
where $\dot{}=4c_4/c_1^2\partial_\tau$, $\mathrm{A}=\zeta(4c_4c_7/c_1^2)^{1/2}$, $\alpha=4c_4c_5/c_1^2$, $\beta=2c_6/c_1\sqrt{|c_7|/c_4}$ and $\Delta=\Delta_2/k_\mathrm{c}^2$ for $k_\mathrm{c}=|c_1|/2c_3$.
We note that $\Delta$ terms are characteristic of a CH coarsening system \cite{Cahn1958} and the polynomial terms appear in SH systems \cite{Cross1993}. For CH-terms to dominate the SH-terms requires the $\tfrac{2}{15}\hat{\mathrm{M}}+3\hat{\mathrm{L}}\ll \tfrac{37}{40},$ where $\mathrm{\hat{L}}=\mathrm{La}^{-2}$ and $\mathrm{\hat{M}}=\mathrm{M}^{-2}$. In high $\hat{\mathrm{M}},\hat{\mathrm{L}}$ regions, the CH terms in the amplitude equation dominate over the SH terms (henceforth called the CH-regime) and vice-versa for low $\hat{\mathrm{M}},\hat{\mathrm{L}}$ regions (the SH-regime). In our experiments, the CH-regime occurs for a concave hemisphere, such as those shown in figure (\ref{fig:patternformation}), and the SH-regime occurs for a convex hemisphere, such as the case shown in figure (\ref{fig:hemispherepf}). Moreover in the SH-regime, weakly-nonlinear stability analysis \cite{Golovin2006} can yield exact regions of the non-dimensional parameters $\mathrm{\hat{L}}$ and $\mathrm{\hat{M}}$ for which different patterns are stable. By comparing the SH-terms of the amplitude equation (\ref{eq:amplitude-equation-doublewell}), i.e. neglecting the CH-terms, with a standard SH system, one finds the stability criteria 

 \begin{alignat}{2}
\text{Hexagonal dots:} &  & -\tfrac{1}{15}\beta^{2}<1-\alpha & <\tfrac{4}{3}\beta^{2}\label{eq:stab1}\\
\text{Mixed dots/Labyrinth:} &  & \tfrac{4}{3}\beta^{2}<1-\alpha & <\tfrac{16}{3}\beta^{2}\label{eq:stab2}\\
\text{Labyrinth:} &  & \tfrac{16}{3}\beta^{2}<1-\alpha\label{eq:stab3}
\end{alignat}
  
where 
 \begin{alignat}{1}
\alpha & =\tfrac{4}{3}\Xi^{2}\hat{\mathrm{M}}, \\
\beta & =\Xi\hat{\mathrm{M}}\bigl(\tfrac{13}{15}\hat{\mathrm{M}}+\tfrac{4}{15}\hat{\mathrm{L}}+\tfrac{56}{225}\bigr)^{-1/2},
\end{alignat}  
for $\Xi^{2}\bigl(\frac{2}{15}\hat{\mathrm{M}}-\hat{\mathrm{L}}\bigr)^{2}=-\tfrac{38}{75}+\tfrac{6}{15}\hat{\mathrm{L}}$. This criteria is visualised in figure (\ref{fig:uwu}a). \changes{In addition, the stability criteria in the SH-regime predicts the existence of two hysteresis cycles. 
With comparison to similar studies on the hysteresis of the SH system \cite{Garcia1992,Kubstrup1996,Stoop2015}, we consider the amplitude of the stationary solution in terms of the parameter $1-\alpha$. The first hysteresis cycle occurs near the transcritical bifurcation point $1-\alpha=0$ and encompasses the subcritical hexagaonal (hysteresis) region $-\tfrac{1}{15}\beta^2<1-\alpha<0$ as shown in figure (\ref{fig:6hysteresis1}a). Within the subcritical hysteresis region, we observe experimentally in Fig. (\ref{fig:hemispherepf}) the possible stable co-existence of both the hexagonal dot state and the no pattern state. For the labyrinth state, we consider the stationary solution of the SH equation with the monoharmonic approximation  \cite{Golovin2006} to yield the amplitude 
\begin{equation}
	\mathcal{A}=\tfrac{2}{\sqrt{3}}\sqrt{1-\alpha},\label{eq:labyamp}
\end{equation}
which is independent of $\beta$. This amplitude solution (\ref{eq:labyamp}) sets up the second hysteresis cycle in the range $\tfrac{4}{3}\beta^{2}<1-\alpha <\tfrac{4}{16}\beta^2$ as shown in figure (\ref{fig:6hysteresis1}b).
 }

\section*{Comparisons with experiments}

Using the amplitude equation (\ref{eq:amp-eqn}), we can identify and graphically quantify the pattern-forming behaviours within the nonlinear system. In particular cases, we can apply the stability criteria in equations (\ref{eq:stab1}-\ref{eq:stab3})  directly to anticipate the dynamics of the pattern formation and verify with experiments using the metric of local fractal dimension. 
In figure (\ref{fig:patternformation}), we have a localised system of a concave thin-film configuration, whereby there is a gradient of film thickness formed by the gravitational drainage flow (see Supplementary Video 2). We observe that the theory closely mimics the experimental data qualitatively in terms of the onset and initial evolution of the dot pattern. 
This suggests that while the amplitude equation (\ref{eq:amp-eqn}) is a leading-order approximation of the system, it nonetheless captures the key behaviour of the system where the diffusional CH-terms dominate over the convectional SH-terms in the amplitude equation (see Supplementary Video 4). We see in (\ref{fig:patternformation}d,h) that the theory starts to deviate from the experimental results as the higher-order behaviours and convection flows due to localised gradients become relevant. To show the finer differences between the experimental image and the simulation result in figures (\ref{fig:patternformation}e-l) in a quantitative manner, we obtain numerically the fractal (Minkowski-Bouligand) dimension (FD) \cite{Theiler1990} of the image divided into $N$ vertical segments, where the FD of a region $\Omega\in\mathbb{R}$ is defined by $\lim_{\delta\rightarrow 0} \log N_\delta/\log (1/\delta)$, for $N_\delta$ is the minimum number of sets with diameter of at least $\delta$ required to cover $\Omega$. Figures (\ref{fig:patternformation}k,l) show a consistent difference in FD between the experimental and simulation results later in the pattern-forming evolution. This quantifies the differences between the full nonlinear system and the leading-order asymptotic model, which captures all but the finest details of the microscale behaviour in the pattern formation dynamics.  It is expected that this difference vanishes upon the consideration of higher-order asymptotics. Another feature of note is that the standard deviation of the FD in experimental images (with the exception of the initial condition) is smaller than that of the simulated results. This suggests that the higher-order asymptotic terms contribute also towards the stability of the FD metric.

A particular case in which the CH- and SH-terms are of a comparative magnitude is shown in figure  (\ref{fig:hemispherepf}) in a convex hemispherical thin-film system.  Under this intermediate CH-SH regime, we observe multiple instability fronts where the pattern formation switches from a configuration of uniform hexagonal dots to one of labyrinth state (see Supplementary Videos 3 and 4). We deduce from this that even though the film thickness is a factor in the onset of the pattern-forming instability, the stability criteria in equations (\ref{eq:stab1}-\ref{eq:stab3}) still governs the state of the instability, i.e. the local pattern formation is not only a function of the film thickness initial condition but also the local values of the Marangoni ($\mathrm{M}$) and Laplace ($\mathrm{La}$) numbers. Assuming the local curvature does not deviate from the initial condition in the particular case as shown in figure (\ref{fig:hemispherepf}a-d), then the Marangoni number is an indicator of the state of the pattern formation. We consider here the Marangoni number calculated from the simulation result in figures (\ref{fig:hemispherepf}i-l). Moreover, a slice of the Marangoni number across the hemisphere from the simulation in figures (\ref{fig:hemispherepf}i-l) shows that the peak values of the $\mathrm{M}$ correspond to regions of labyrinth behaviour, whereas trough values denote hexagonal states. Physically, this suggests that a greater local variation of the surface tension neccessarily leads to a more disordered and symmetry-breaking pattern formation, whereas a smaller local perturbation in the local surface tension creates the more orderly and symmetrical hexagonal state of pattern formation. 


\begin{figure*}
  \includegraphics[width=18cm]{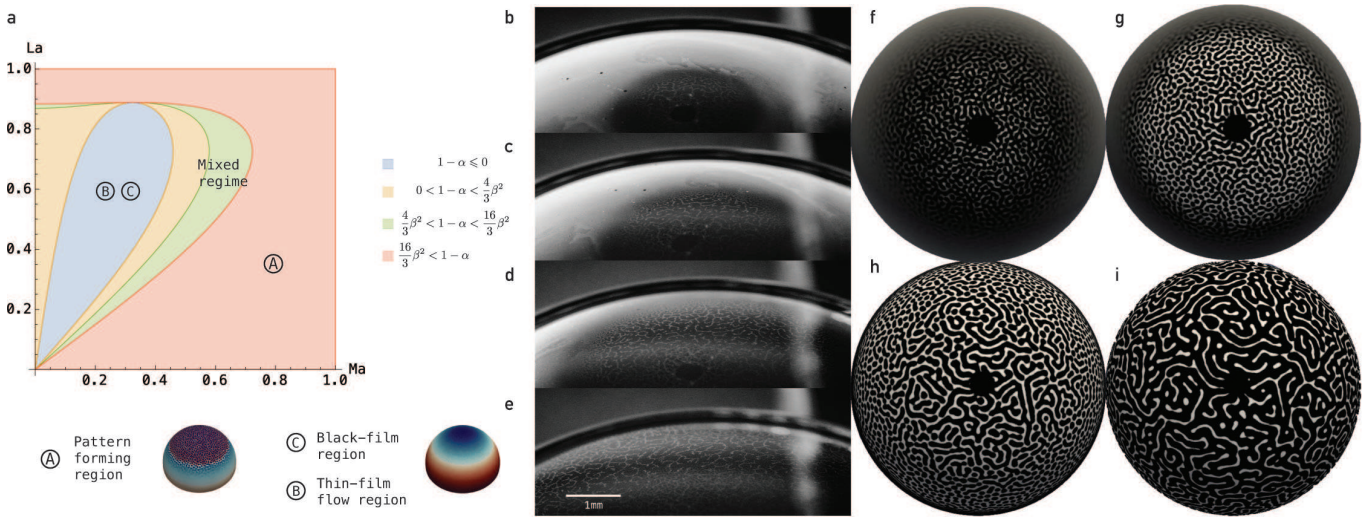}
  \caption{$\mathbf{a}$. Plot of the stability diagram, where (A) denotes regions of pattern formation, and (B) and (C) denote the thin-film flow and the black-film regions, respectively. $\mathbf{b-e}$. Experimental images of the labyrinth state on a convex hemispherical thin-film system. $\mathbf{f-i}$. FEM simulation of the amplitude equation (\ref{eq:amplitude-equation-doublewell}) in region (A).}
  \label{fig:uwu}
\end{figure*}

Another case of the convex hemispherical thin-film system under the intermediate CH-SH regime, now with pattern formation predominantly in the labyrinth state, is shown in figure (\ref{fig:uwu}b-e) and Supplementary Videos 3 and 4. More particularly, we observe in both figures (\ref{fig:hemispherepf}a-b) and (\ref{fig:uwu}b-e) the appearance of a circular ``black hole" region at the topmost part of the hemisphere, corresponding to the thinnest portion of the whole system.
We denote this the black-film region C in figure (\ref{fig:uwu}a), wherein the molecules at the interface are in a close packing configuration \cite{Exerowa1981}. Locally, this close packing correlates to a fully-adsorbed interface which results in the homogeneous condition that there is negligible Marangoni gradient, i.e. $\mathrm{M}\ll 1$. In terms of the amplitude equation, the black-film region C admits a U-shaped quadratic energy functional in the CH-terms of the amplitude equation (\ref{eq:amp-eqn}), which can be linearised to the Helmholtz form $(\Delta_{2}+\Omega^2)\zeta=0,$ where the film thickness $\zeta$ can be interpreted in this case as the amplitude of time-harmonic solutions to the wave equation $\eta_{tt}=\Delta \eta$,  with frequency $\Omega$ given by $\Omega^2 = c_5/c_1$ for $\eta = \zeta \exp(-i\Omega t)$. The boundary condition is given by the thickness-wavelength relation in equation (\ref{eq:th-wa-relation}) that $\lim_{\zeta\rightarrow \zeta_\mathrm{c}}\eta(r_\mathrm{c},\theta,t)= 0$, where $\zeta_\mathrm{c}\sim (\lambda_\mathrm{c}^{\mathrm{w}})^{3/2}$ and $r_\mathrm{c}$ is the radius from the apex where the film has thickness $\zeta_\mathrm{c}$. The simulation in figures (\ref{fig:uwu}f-i) takes into account this black-hole region by imposing an additional Gaussian decay initial condition on the top-most spherical cap with radius $r_\epsilon\ll 1$. 
As $\zeta\rightarrow \zeta_\mathrm{c}^+$, $\mathrm{M}=O(1)$, the patterns correspond to the peaks of the Marangoni number plot, also referred to as the pattern-forming region A in figure (\ref{fig:hemispherepf}k) and (\ref{fig:uwu}a), and evolve according to the amplitude equation (\ref{eq:amp-eqn}) with a W-shaped quartic energy potential $f(\mathrm{A})$. Lastly, in the thin-film region B the dominant dynamics is the thin-film fluid flow rather than the pattern-forming instability and this is shown as $\mathrm{M}$ reaches a trough in figure (\ref{fig:hemispherepf}k). Opposite to the black-film region C, the local Marangoni gradient is negligible due to a lack of coverage of surfactant on the interface as the fluid flow transports surface materials along the interface. Correspondingly in figures (\ref{fig:hemispherepf}m-p), we observe this interfacial transport process congregates the surfactants near the top of the hemisphere as time increases, reinforcing the idea of a two-regime dynamics, where the surfactant-concentrated region exhibits quasi-elastic \cite{Stoop2015} patter-forming behaviours whilst thin-film fluid flow is the dominant dynamics for regions where surfactant concentration is low.

\section*{Conclusions and outlook}

We demonstrated that an effective field theory derived systematically using asymptotics provides a good quasi-quantitative description of the surface pattern formation in non-planar viscous thin liquid films in the presence of a surfactant solution. Systems which exhibit similar pattern forming capabilities can be found across different lengthscales but the underlying principle of symmetry-breaking in the field equation is a common mechanism by which patterns form on the surface of a homogeneous film, and the presence of surfactants along with geometric curvature provides inputs and enables a more systematic classification of the overall wrinkling process in a dynamic situation. Thus, the proposed analytical approach can help us to recognise more quantitatively pattern formation processes beyond a static situation and is a significant step towards characterising completely the complex dynamic behaviour of the thin film near the process of rupture to the leading order.    

An interesting result of this analytical work is that the thin-film naturally yields a power-relation relation between film thickness and the instability wavelength $\lambda$ through the requirement that local fluid velocities are finite. Moreover, the proximity of the critical instability wavelength $\lambda_c$ to critical damping wavelength of the capillary wave, $\lambda_{\mathrm{c}}^{\mathrm{w}}$, as observed experimentally with interferometry, together with previous analytical work on capillary waves in the presence of surfactants \cite{Shen2018} prompts a conjecture between the two quantities. We expect the capillary wave theory together with a more sensitive analysis of transient film compositions to be instrumental in understanding film behaviour and the nucleation of black spots \cite{Derjaguin1994} near the rupture process and that the capillary wave theory provides a  useful vehicle through which we can understand the damping behaviours and the roughness of the interface in the neighbourhood of the onset of the pattern-forming instability. We are encouraged by recent results in pattern formation for stationary elastic spherical objects \cite{Stoop2015} and we anticipate that our analytical theory leads to further experimental work in the measurement of the relevant quantities in transient and fluidic pattern-forming phenomena.

\section*{Algorithm}

The simulation of the amplitude equation in equation (\ref{eq:amplitude-equation-doublewell}) is non-trivial. While the double-well potential is reminiscent of the Cahn-Hilliard (CH) equation, the polynomial terms are a feature of the Swift-Hohenberg equation, coupled with the fourth-order bi-Laplacian operator. This presents challenges within the finite element method. To employ a Galerkin-method requires either a piecewise smooth and globally $C^1$-continuous element for a basis function or a mixed formulation which bypasses the $C^1$-continuous requirement by introducing an auxiliary field to recast the fourth-order equation of motion into two coupled second-order equations. Here we used the mixed formulation which has been shown, for CH-class equations \cite{Zhang2013,Kaessmair2016}, to be less computationally expensive with a comparable accuracy to $C^1$-continuous methods. To realise the finite element formulation, the solution of a system
state is given via the interpolation function $A(\mathbf{x})=\sum_{i=1}^{N}A_{i}N_{i}(\mathbf{x}),$
where $N_{i}$ are the finite-element basis and $A_{i}$ are the coefficient
for each triangular elements with index $i=1,\ldots,N$. The mesh used for the Galerkin projection is a triangulation of the interface with $97,000$ elements for the (convex) hemispherical case and $87,000$ elements for the (quasi-)planar concave case. The FEM package FENiCs \cite{Alnaes2015} is used for the calculation, whereby integrals over the manifold mesh are reduced to integrals over the surface facets of a mesh \cite{Rognes2013}. The Crank-Nicolson method is used for the discretisation in time and the Newton-Krylov solvers based on PETSc's SNES module is used with the discretisations in space and time solved using the general minimal residual method (GMRES). Each iteration is solved to a relative tolerance of $10^{-6}$. The solution process scales well with multiple cores using the MPI routine.

\section*{Experiment}
The concave hemispherical soap films were created and stabilised on a bath of diluted liquid detergent solution (Fairy liquid, Procter \& Gamble) to allow a natural and undisturbed formation of a spherical geometry. The effective radius of the concave hemispherical membranes were chosen to be $5\mathrm{cm}\leqslant R_{\text{concave}}\leqslant 8\mathrm{cm}$ such that we can neglect any edge and contact line effects. The convex thin film was formed on a apparatus with a spherical gap. The radius was fixed at $R_{\text{convex}}=2.5\,\text{cm}$. The experimental images were taken using a Nikon DSLR with a 105mm macro lens at 1:1 magnification. More details of the experiment can be found in the Supplementary Material. 

\section*{Acknowledgements}
This work was supported by the Shell University Technology Centre for fuels and lubricants and the Engineering and Physical Sciences Research Council (EPSRC) through grants EP/M021556/1 and EP/N025954/1. In addition, the authors acknowledge the fruitful discussions with Prof. S. Kalliadasis regarding the hysteresis of the pattern formation process and the conditions of the asymptotic expansion. 

\vspace{1cm}

\section*{Author contributions}
L.S., F.D. and D.D. conceived the ideas and developed the research strategy. L.S. developed the theory, the experimental setup and performed the analytical and numerical calculations. All authors discussed the results and contributed to writing the paper.

\bibliographystyle{apsrev4-1} 
\bibliography{Bib.bib}


\end{document}